\documentclass[USenglish,twocolumn]{article}

\usepackage{graphicx,import}
\usepackage{amssymb, cuted}
\usepackage{floatrow} 
\usepackage{subfig}
\usepackage{lipsum}
\usepackage{anyfontsize}
\usepackage{cuted}
\usepackage[export]{adjustbox}
\usepackage{bibentry}
\usepackage{ulem}
\usepackage[dvipsnames,table]{xcolor}
\usepackage{tikz}
\usepackage[ruled,longend]{algorithm2e}
\usetikzlibrary{patterns,arrows,decorations.pathreplacing}

\newcommand{\norm}[1]{\left\lVert#1\right\rVert}

\usepackage{cite}
\usepackage{xcolor}

\usepackage[utf8]{inputenc}				
\usepackage[big,online]{dgruyter}	
\usepackage{lmodern} 
\usepackage{microtype}
\usepackage[numbers,square,sort&compress]{natbib}

\theoremstyle{dgthm}

\theoremstyle{dgdef}

\begin{document}

  \startpage{1}
  \aop

\title{Master memory function for delay-based reservoir computers with single-variable dynamics}
\runningtitle{Master memory function for delay-based reservoir computers with single-variable dynamics}

\author*[1]{Felix Köster}
\author[2]{Serhiy Yanchuk}
\author[1]{Kathy Lüdge} 
\runningauthor{F.~Köster et al.}
\affil[1]{\protect\raggedright 
Institut
for Theoretical Physics, Technische Universität Berlin, Berlin,
10559 Germany: f.koester@tu-berlin.de}
\affil[2]{\protect\raggedright 
Institut
for Mathematics, Technische Universität Berlin, Berlin,
10559 Germany}
	
	
\abstract{We show that many delay-based reservoir computers considered in the literature can be characterized by a universal master memory function (MMF). 
	Once computed for two independent parameters, this function provides linear memory capacity for any delay-based single-variable reservoir with small inputs. Moreover, we propose an analytical description of the MMF that enables its efficient and fast computation.
	Our approach can be applied not only to reservoirs governed by known dynamical rules such as Mackey-Glass or Ikeda-like systems, but also to reservoirs whose dynamical model is not available. We also present results comparing the performance of the reservoir computer and the memory capacity given by the MMF.}

\keywords{Reservoir Computing, Nonlinear Dyanmics, Machine Learning}

\maketitle

\section{Introduction} 

Reservoir computing is a neuromorphic inspired machine learning paradigm, which enables high-speed training of recurrent neural networks and is capable of solving highly complex time-dependent tasks.
First proposed by Jaeger \cite{JAE01} and inspired by the human brain \cite{MAA02}, it utilizes the inherent computational capabilities of dynamical systems.  
Very recently, the universal approximation property has also been shown for a wide range of reservoir computers, which solidifies the concept as a broadly applicable scheme \cite{GON20}.
Bollt pointed out a connection between reservoir computers and VAR (vector autoregressive) and nonlinear VAR  machines, which may be one of the reasons behind the surprising efficiency of reservoir computers for time-dependent tasks \cite{BOL21,Gauthier2021a}.
Many different realizations \cite{BAU15, KEU17, SCA16, ARG17, ARG18, AMI19, PAT18, PAT18a, CUN19, VAN14, FER03, ANT16, DOC09} have shown the relevance of reservoir computing to practical applications, while analytical and numerical analyses \cite{GAL18a, GAL19, GRI14, GRI15} help in building understanding of its working principles and improve its performance.
Motivated by fast inference  and low energy consumption,  optoelectronic and optical hardware implementations of reservoir computers are often realized  \cite{LAR12, PAQ12, BRU13a, VIN15, NGU17, ROE18a, ROE20, BUE18a, BUE17, NAK16} indicating a high future potential of such processing unit.

\begin{figure}%
	\centering
	\includegraphics[width=\textwidth]{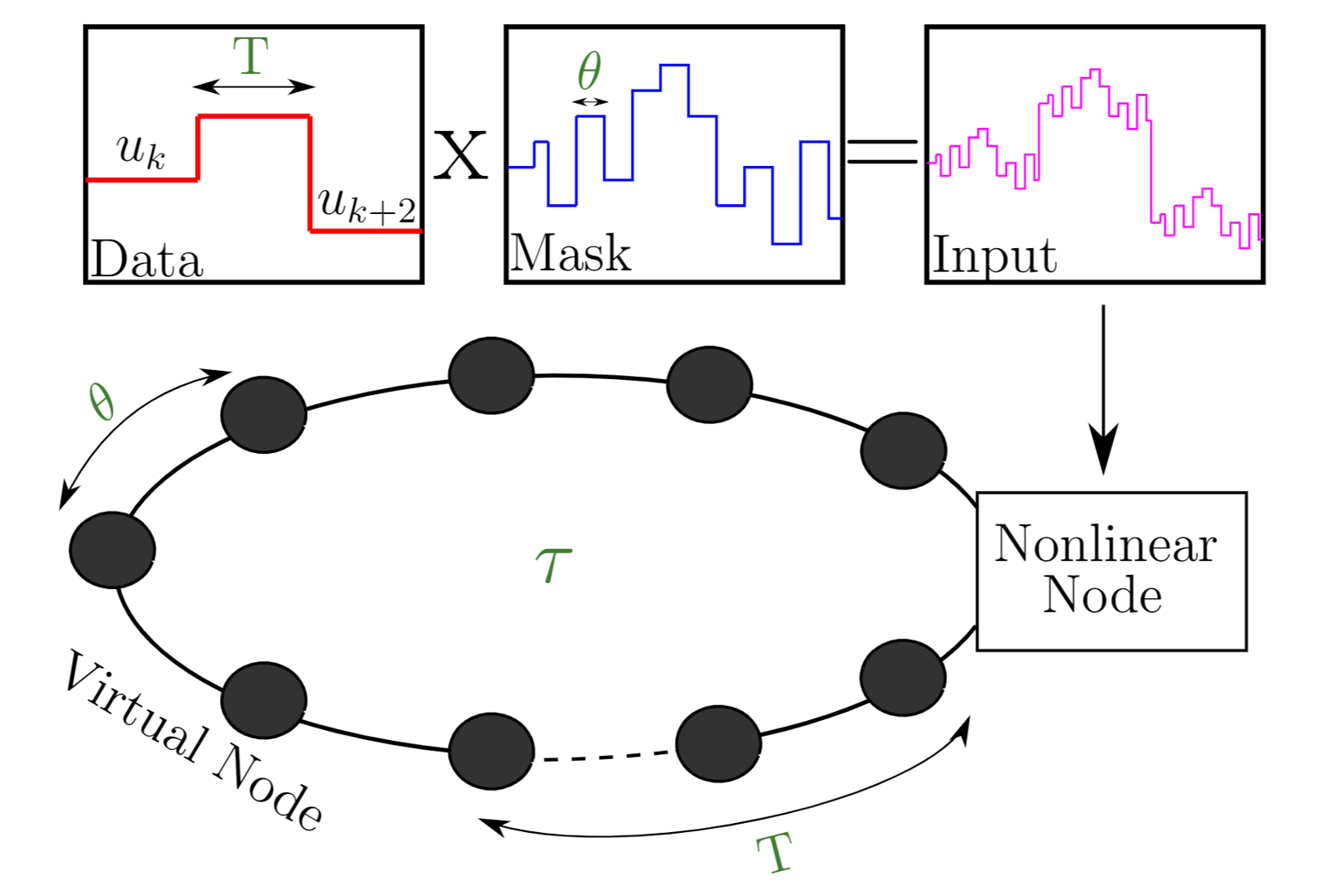}
	\caption[schematic]{Scheme of delay-based reservoir computing. Important timescales are marked in green: $T$ input cycle, $\tau$ delay, $\theta$ virtual node seperation time.}
	\label{fig:sketch}
\end{figure}

Originally, reservoir computing is performed with a network of nonlinear nodes, which projects the input information into a high dimensional phase space, allowing a linear regression to linearly separate features \cite{JAE01}.
In time-delayed reservoir computing, a single dynamical node with delayed feedback is employed as a reservoir instead of the network \cite{APP11}. 
The time-multiplexing procedure allows for such a single-element system to implement a recurrent ring network \cite{APP11,HAR17a,STE21}, see 
Fig.~\ref{fig:sketch}.
The absence of the need for a large number of nonlinear elements significantly reduces the complexity of the reservoir hardware implementation.
Existing experimental and numerical realizations show promising results in solving time-dependent tasks, such as speech recognition, time-series predictions \cite{LAR17,BUE17,KUR18,ORT17a,DIO18,BRU18a,CHE19c,HOU18,SUG20} or equalization tasks on nonlinearly distorted signals \cite{ARG20}.
For a general overview, we refer to \cite{BRU19,SAN17a,TAN18a}.

Often reservoirs are optimized to a specific task by hyperparameter tuning, which defeats the purpose of reservoir computing as a fast trainable machine learning scheme.
Dambre et al. \cite{DAM12} introduced a task-independent quantification of a reservoir computer, 
building on the memory capacity notion already introduced in \cite{JAE01} 
whereas a high memory capacity pinpoints to generally well-performing reservoirs.

In this paper, we provide an analytical tool for finding promising reservoir setups by introducing a master memory function (MMF) for delay-based reservoir computing with a small input. The MMF allows for fast computable predictions of the linear memory capacity and it indicates that linear memory capacity of reservoirs is similar for systems with similar linearizations.

The main idea behind our method can be outlined as follows. Consider a delay-based reservoir described by a general nonlinear system $\dot s (t) = f(s(t),s(t-\tau),I(t))$, where $I(t)$ is an input signal, which is ''small'' in a certain sense, and $s(t)$ determines the state of the reservoir. The response $s(t)$ of the reservoir must be independent (at least to some extend) on its initial state, the property known as echo state. Such a situation occurs when the 
reservoir is operating near an equilibrium state $s^*$ that is stable in the absence of the input signal. 
Therefore, all reservoir dynamics takes place in a neighborhood of this equilibrium and as a result, the reservoir linearization  $\delta \dot  s (t) = a\delta s(t) + b\delta s(t-\tau) + cI(t)$ approximates these dynamics. Here $\delta s$ is the deviation from the equilibrium. In the considered case of the single-variable reservoir, the scalar parameters $a$ and $b$ are the only determining quantities. The relatively simple form of the linearized system allows us to obtain an analytical expression for the linear memory capacity, which depends on the parameters $a$ and $b$ and thus parametrically determines the linear memory capacity of any reservoir with the above  properties. We call the obtained function MMF due to its universal features, i.e. different reservoir computing setups, which possess the same linearizations, yield the same linear memory capacity given by MMF.

The paper is structured as follows. First, we will briefly revise the concept of time-delay-based reservoir computing and the concept of linear memory capacity.
We will then present our main analytical result while additionally presenting an example code for an efficient evaluation of the obtained expression;  the derivation is given in the appendix.
Finally, comparisons of  numerically simulated reservoir computer performance with the semianalytical approach are provided.
We also show in Sec.~\ref{sec:ssr} the applications of our results to reservoirs with unknown dynamical model, where the parameters $a$ and $b$ are evaluated using the system response to an external stimuli.

\section{Time-Delay based Reservoir Computing}
\label{sec:methods}

Reservoir Computing utilizes the intrinsic abilities of dynamical systems to project the input information into a high dimensional phase space \cite{JAE01}.
By linearly combining the responses of the dynamical reservoir to inputs, a specific task is approximated.
In the classical reservoir computing scheme, often, a so-called echo state network is used by feeding the input into a spatially extended network of nonlinear nodes.
Linear regression is then applied to minimize the Euclidean distance between the output and a target.
This approach is particularly resourceful for time-dependent tasks because the dynamical system which forms the reservoir acts as a memory kernel.

In the time-delay-based reservoir computing scheme \cite{APP11}, the spatially extended network is replaced by a single nonlinear node with a time-delayed feedback loop.
The time-multiplexing procedure with a periodic mask function $g$  is applied to
translate the input data to a temporal input signal. Similarly, the time-multiplexing procedure translates the single temporal high-dimensional reservoir response to the 
spatio-temporal responses of virtual nodes. 
The virtual nodes play the same role as the spatial nodes in echo state networks. 

A sketch of the delay-based reservoir computing setup is shown in Fig. \ref{fig:sketch}.
In the following, we will give a short overview of the quantities and notations used in this paper.
We also refer to our previous works \cite{KOE21,KOE20a,STE20} for a detailed explanation of how the reservoir setup is operated and task-independent memory capacities are computed.

\begin{figure}%
	\centering
	\hspace*{-0.4cm} 
	\includegraphics[width=1.1\textwidth]{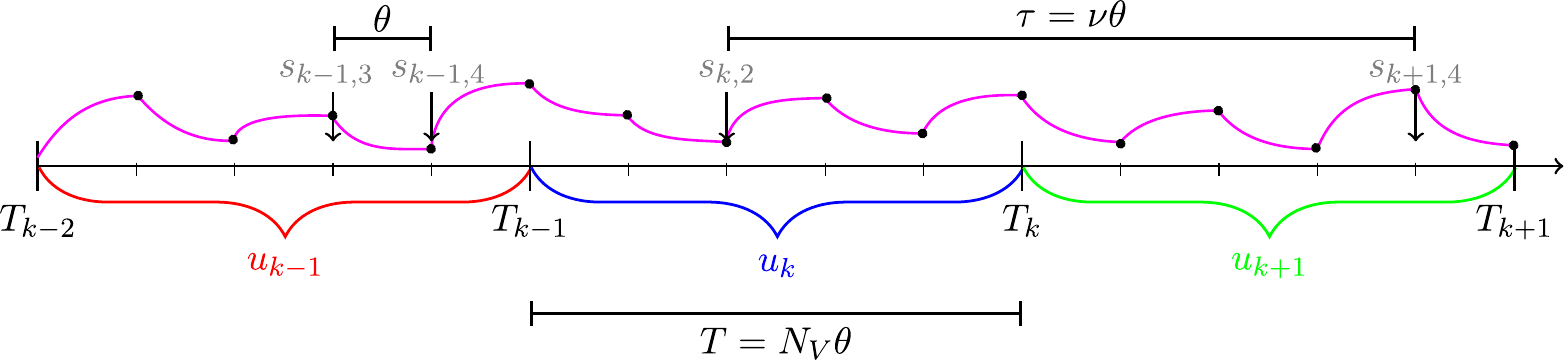}
	\caption[Timeline]{Exemplary timeline sketch for time-delay based reservoir computing. Three input intervals of length $T$ are shown for the inputs $u_{k-1}$, $u_k$, $u_{k+1}$ in red, blue and green respectively. The delay time $\tau$ in multiplicatives of the $\theta$-intervals is $\nu=7$. The number of virtual nodes in this example is $N_V=5$, thus $\tau > T$. Four system states are indicated in grey ($s_{k-1,3}$, $s_{k-1,4}$, $s_{k,2}$ and $s_{k+1,4}$), where $s_{k,2}$ influences $s_{k+1,4}$ directly via the delay time $\tau$. The pink line indicates an example trajectory, with black dots showing the measured system states i.e., the virtual nodes. }
	\label{fig:example_timeline}
\end{figure}%

\begin{figure}
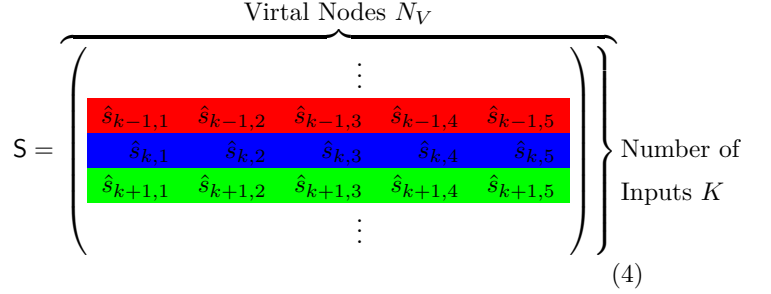

	\centering
	\begin{align}
		\mathsf{S} =  \overbrace{ \left.\left(  \begin{array}{rrrrr}
				&  & \vdots & & \\ 
				\rowcolor{red}
				\hat s_{k-1,1} & \hat  s_{k-1,2} & \hat   s_{k-1,3} &  \hat  s_{k-1,4} & \hat    s_{k-1,5} \\
				\rowcolor{blue}
				\hat  s_{k,1} & \hat  s_{k,2} & \hat  s_{k,3} & \hat  s_{k,4} & \hat   s_{k,5} \\
				\rowcolor{green}
				\hat  s_{k+1,1} & \hat  s_{k+1,2} & \hat  s_{k+1,3} &  \hat  s_{k+1,4} & \hat   s_{k+1,5} \\
				&  & \vdots & & \\ 
			\end{array} \right)\right\} }^{\text{\normalsize{Virtal Nodes} } {\displaystyle N_V}} \begin{aligned}[t]&\text{Number of}\\
			&\text{Inputs $K$}\end{aligned}
	\end{align}
	\caption{State matrix $\mathsf{S}$ corresponding to the timeline shown in Fig. \ref{fig:example_timeline} with $N_V=5$.
	}
	\label{fig:example_matrix}
\end{figure}

Let us briefly remind the main ingredients of the time-multiplexed reservoir computing scheme \cite{APP11,KOE20a,STE20, KOE21}.
We apply an input vector $\mathbf{u}\in \mathbb{R}^K$ componentwise at times $t\in [t_{k-1},t_k)$, $t_k=kT$, $k=1,\dots,K$, $K$ being the number of sample points.
The administration time for different  inputs $t_{k+1} - t_k = T$ is the same and it is called the clock cycle $T$.
To achieve a high dimensional response to the same input, a $T$-periodic mask function $g$ multiplies the input and the resulting signal enters the system (see Fig. \ref{fig:sketch} and Fig. \ref{fig:example_timeline}).
The mask $g$ is a piecewise-constant function on $N_V$ intervals, each of length $\theta=T/N_V$ corresponding to $N_V$ virtual nodes.
The values of the mask function $g$ play the same role as the input weights in spatially extended reservoirs, with the difference that time-multiplexing distributes the weights over time.
The responses of the reservoir are collected in the state matrix $\mathsf{S} \in \mathbb{R}^K \times \mathbb{R}^{N_V}$, see
Fig. \ref{fig:example_matrix}.
The elements of the state matrix are 
$\left[ \mathsf{S}\right]_{km}= \hat s(kT+m\theta)$ with $m=1,\dots,N_V$, and $k=1,\dots,K$, where $\hat s((kT+m\theta))\in\mathbb{R}$ is the state of the dynamical element of the reservoir at time $(kT+m\theta)$ shifted by the mean over all clock cycles $\hat s(kT+m\theta) = s(kT+m\theta) - \left\langle  s(\cdot \, T+m\theta) \right\rangle$, see \cite{DAM12}. The average $\left\langle  s(\cdot \, T+m\theta) \right\rangle$ can be understood as the averaging over the row elements. For example, for an experimental or numerical realization of the reservoir with a semiconductor laser, $s(t)$ could be the laser intensity.

A linear combination of the state matrix is given by $\mathsf{S} \mathbf{w}$, where $\mathbf{w}\in \mathbf{R}^{N_V}$ is a vector of weights. Such a combination is trained by ridge regression, i.e., the least square approximation to some target vector $\mathbf{\hat{y}}$
$$
\arg\min_\mathbf{w} \left[ \| \mathsf{S} \mathbf{w} - \mathbf{\hat{y}} \|_2^2 + \lambda \| \mathbf{w} \|_2^2\right],
$$
where $\|\cdot\|_2$ is the Euclidean norm, and $ \lambda$ is a Tikhonov regularization parameter.  The solution to this problem is 
\begin{align}
	\mathbf{w} = (\mathsf{S}^T \mathsf{S} + \lambda \mathrm{I})^{-1} \mathsf{S}^T \mathbf{\hat{y}}.
\end{align}
In the case of invertible $\mathsf{S}^T\mathsf{S}$, the matrix $(\mathsf{S}^T \mathsf{S} )^{-1} \mathsf{S}^T$ is the Moore-Penrose pseudoinverse. 
We set $\lambda=10^{-6} \cdot \max_{s} (\mathsf{S})$, where $\max_{s} (\mathsf{S})$ is the largest state response in the state matrix $\mathsf{S}$. 
To quantify the system's performance, we use the capacity (see \cite{DAM12,KOE21}) $\text{C}_{\mathbf{\hat{y}}}$ to approximate a specific task which is given by
\begin{align}
	\text{C}_{\mathbf{\hat{y}}} = 1 - \text{NRMSE},
	\label{eq:2}
\end{align}
where NRMSE is the normalized root mean square error  between the approximation $\mathbf{y}=\mathsf{S} \mathbf{w}$ and the target $\mathbf{\hat{y}}$
\begin{align}
	\text{NRMSE} = 
	\left( 
	\frac{\sum\limits_{k=1}^{K}(\hat{y}_{k} - y_{k})^2}
	{N \cdot \mathrm{var}(\mathbf{\hat{y}})} 
	\right)^{1/2}
	,
\end{align}
where $\mathrm{var}(\mathbf{\hat{y}})$ is the variance of the target values $\mathbf{\hat{y}}=(\hat{y}_1,\ldots,\hat{y}_K)$.

\section{Reservoir Computation Quantification }
Here we introduce the linear memory capacity as a quantitative measure for the memory kernel of a dynamical system.

\subsection{Memory Capacity}
The central task-independent quantification was introduced by Jaeger in \cite{JAE01} and refined by Dambre et al. in \cite{DAM12}, which yields that the computational capability of a reservoir system can be quantified via an orthonormal set of basis functions on a sequence of inputs. 
Here we give a recap of the quantities introduced in \cite{KOE20a, KOE21} and focus on the linear memory capacity. \\
In particular, the capacity to fulfill a specific task is given by 
\begin{align} 
	\text{C}_{\mathbf{\hat{y}}} = \frac{\mathbf{\hat{y}}^T \mathsf{S} (\mathsf{S}^T\mathsf{S} + \lambda \mathrm{I})^{-1} \mathsf{S}^T \mathbf{\hat{y}}}{\norm{\mathbf{\hat{y}}}^2}
	= 
	\frac{\mathbf{\hat{y}}^T \mathbf{y}}
	{\norm{\mathbf{\hat{y}}}^2},
	\label{eq:mpsi_mem_capacity}
\end{align}
which can be derived from Eq.\,(\ref{eq:2}) (see \cite{DAM12,KOE21}).
The capacity equals $1$ if the reservoir computer computes the task perfectly, thus $\mathbf{y}=\mathbf{\hat{y}}$; and it equals 
$C=0$ if the prediction is not correlated with the target. In between $0$ and $1$ if it is partially capable of fulfilling the task. 
To quantify the systems capability for approximating linear recalls of inputs, an input sequence $\{u\} = \{ u_{-K}, \dots, u_{-3}, u_{-2}, u_{-1}\}$ is applied, where $u_k$ are uniformly distributed random numbers, independent and identically drawn in $[-1,1]$.
With the input sequence $\{u\}$ of random numbers, the  reservoir response is collected in the state matrix $\mathsf{S}$. 

To describe a linear recall task of $l$ steps into the past, the
target vector $\mathbf{\hat{y}}$ is defined as
\begin{align}
	\mathbf{\hat{y}}_l = 
	\{ \dots, u_{-3-l}, u_{-2-l}, u_{-1-l}\},
	\label{eq:LP_construction}
\end{align}
which is the linear recall $l$ steps into the past. 
Formally, one considers an infinitely long sequence $K\to\infty$. To approximate it numerically, we use $K=75000$.

The \textit{linear memory capacity} $\text{MC}$ is defined as the sum over the capacities of all possible linear recall tasks
\begin{align}
	\text{MC} = \sum_{l=0}^{\infty} C_l,
	\label{eq:tasks-d}
\end{align}
where $C_l = C_{\mathbf{\hat{y}_l}}$ is the capacity of the $l$-th recall into the past.
This quantification is task independent and thus implications for specific applications cannot be given. 
Different tasks may need different specific capacities.
The measure $\text{MC}$ thus only gives a hint for well-performing reservoirs in the context of using the full scope of the given reservoirs, rather than a direct task-specific estimate.
We have to point out, that the linear-nonlinear trade off is a well-known effect \cite{DAM12}, thus a system with high linear memory capacity can yield a low nonlinear transformation capability.
Nevertheless, we believe predicting a well-performing linear memory kernel reservoir is beneficial for a general reservoir computer setup, as higher nonlinear memory transformation can be utilized by adding additional reservoir systems with increased perturbations.

\subsection{Reservoir Systems}

As one example system, we use a Stuart-Landau oscillator with delayed feedback:
\begin{align}
	\frac{ds(t)}{dt} = (p_{SL} + \eta I(t) + \gamma s(t)^2)s(t) + \kappa s(t - \tau) \label{eq:ST}
\end{align}
Here, $s(t)$ describes the real-valued amplitude of the system, $\kappa$ is the feedback strength, $\tau$ the delay time,
$p_{SL}$ is a parameter determining the dynamics, and $\eta$ the input strength of the information fed into the system. 

For $p_{SL}+\kappa>0$ and $\eta=0$, system (\ref{eq:ST})
has only the trivial equilibrium $s=0$, and for $p_{SL}+\kappa<0$, additionally, the nontrivial equilibria exist $(s^{*})^{2} = -(p_{SL} +\kappa)/\gamma$, which appear in a pitchfork bifurcation at $p_{SL}+\kappa=0$. The linearization at the nontrivial equilibria (taking into account the input term) reads
\begin{align}
	\frac{\delta s(t)}{dt} = a \delta s(t) + b \delta s(t - \tau) + c I(t),
	\label{eq:linearization}
\end{align}
where $a = -2p_{SL} - 3\kappa = p_{SL} + 3\gamma (A^{*})^{2}$, $b=\kappa$, and $c=\eta s^*$.

As the second example, we use the Mackey-Glass system
\begin{align}
	\frac{ds(t)}{dt} = (p_{MG} + \eta I(t))s(t) + \frac{\alpha s(t - \tau)}{1 + s^{p}(t - \tau)},
	\label{eq:MG}
\end{align}
where $s(t)$ is the dynamical variable, $s(t - \tau)$ is the delayed variable, and $p_{MG}$, $p$, and $\alpha$ are control parameters.
The reservoir input is fed into the system via the  term $\eta I(t)$.
We set $p=1$, for which the system possesses a
stable nontrivial equilibrium 
$s^{*} = -(p_{MG} + \alpha)/p_{MG}$ (for $\eta=0$). The corresponding linearization at this equlibrium is  Eq.~\eqref{eq:linearization} with $a = p_{MG}$, $b=\alpha/(1 + s^{*})^2$, and $c=\eta s^*$.

\section{Results}
\label{Results}


\subsection{Analytic description of memory capacity}

From Eq. (\ref{eq:mpsi_mem_capacity}), we see that the capacity to approximate a specific input is given by the inverse of the covariance matrix $\left(\mathsf{S}^T\mathsf{S}\right)^{-1}$ (corrected by $\lambda \mathrm{I}$), also called the concentration matrix, and the matrix multiplication of the state matrix and the target $\mathsf{S}^T \mathbf{\hat{y}}$.
Thus, it is necessary to derive the state matrix $\mathsf{S}$ from the responses to the small perturbations of the system.
This has already been done for 1-dimensional reservoirs with $\tau=T$ by an Euler step scheme \cite{GRI15}, and for 1-dimensional reservoirs with $\tau \neq T$ for specific differential equations \cite{STE20}.
We would like to extend this knowledge by analyzing arbitrary systems and $\tau \neq T$.
We assume the virtual node distance $\theta$ to be small and $\tau=\nu \theta$, with $\nu \in \mathbb{N}^{+}$.
We also assume the operation point of the reservoir to be a stable equilibrium.
We will exemplarily validate our analysis on the two 1-dimensional nonlinear reservoirs given by Eqs. (\ref{eq:ST}) and (\ref{eq:MG}).

Our main result is the modified state matrix $\tilde{\mathsf{S}}$, that we can use to determine the MC while we can calculate it solely from the linearized system. 
The entries $\left[ \tilde{\mathsf{S}} \right]_{jn}$ are given by 
\begin{align}
	\left[ \tilde{\mathsf{S}} \right]_{jn} = 
	\frac{\gamma}{\sqrt{3}}
	\sum_{n + N_{V} j \le i+k\nu}^{i+k\nu < n + N_{V} (j+1)}
	\binom{k+i}{i}p^{i}m^{k}
	w_{\left(n-i-k\nu\right)\,\text{mod}\,N_{V}},
	\label{eq:xnl}
\end{align}

where $p=e^{a\theta}$,  $m=-(b/a)(1-p)$, $\gamma = - (c/a) (1-p)$, the parameters $a$ and $b$, and $c$ are given by the linearization \eqref{eq:linearization}, $w_{i}$ are the weights of the time-multiplexing. The index
$j$ corresponds to the $j$-th clock cycle, and $n$ to the $n$-th virtual node. 
The rows of the modified state matrix contain entries in the statistical direction of the $l$-th shifted input.
As we show in App. \ref{app:A}, the covariance of the modified state matrix approximates the original state matrix
\begin{align}
	\tilde{\mathsf{S}}^T \tilde{\mathsf{S}} = 
	\mathsf{S}^T\mathsf{S}.
	\label{eq:S_tilde_S_connection}
\end{align}
Moreover, we also show that the full linear memory capacity can be calculated by using solely the modified state matrix
\begin{align}
	\label{eq:MC-11}
	\text{MC} \approx \text{MC}_{\text{MMF}} = \text{tr}\left(\tilde{\mathsf{S}}^T \left(\tilde{\mathsf{S}}^T \tilde{\mathsf{S}}  + \lambda \mathrm{I} \right)^{-1} \tilde{\mathsf{S}}\right),
\end{align}
and the capacity of the $l$-th recall is given by
\begin{align}
	C_l \approx \tilde{\mathsf{S}}^T_l \left(\tilde{\mathsf{S}}^T \tilde{\mathsf{S}} + \lambda \mathrm{I}\right)^{-1} \tilde{\mathsf{S}}_l,
	\label{eq:Cl}
\end{align}
where $\mathsf{\tilde S}_l$ is the $l$-th row of $\mathsf{\tilde S}$.
Details of the derivations can be found in App. \ref{app:A}.

We call the memory capacity given by Eq.~\eqref{eq:MC-11} the Master Memory Function (MMF). For given parameters of the linearization $a$, $b$, and $c$, as well as the mask coefficients $w_j$, this function can be evaluated in a much more efficient way than the direct evaluation of the linear memory capacity via a stepwise integration of the differential equation.
A speed comparison is given in App. \ref{app:B}.
The new approach does not require calculating the reservoir, and it does not involve the input sequence $u_{k}$. 

\subsection{Efficient numerical evaluation of the memory capacity and the modified state matrix}

The obtained approximations of the 
modified state matrix  (\ref{eq:xnl}) and memory capacity function \eqref{eq:MC-11} allow for efficient numerical evaluation. For this, we propose the following scheme, which we also show as pseudocode in Alg. \ref{alg:s_nl}.

First, we iterate over all entries of the modified pascal's triangle given in Fig. \ref{fig:pascals}, which can be done by two nested loops $k$, $i$.
We do this until all entries in a row $q$ are below a given threshold $\epsilon$ for $i+k = q$ for $i,k \in \mathbb{N}$ (see Fig.\,\ref{fig:pascals}).
The threshold $\epsilon$ ensures that we cut unnecessary terms smaller than the regularisation parameter $\lambda$.
A third loop $n$ goes over all virtual nodes $N_V$ adding the result $\binom{k+i}{i} p^i m^k$ multiplied with the corresponding weight $w_{n+i+k\nu \mod N_V}$ to all corresponding entries $\tilde{s}_{\lfloor(n+i+k\nu)/N_V \rfloor,n}$, that thus lie in the same input interval $l$.
See App. \ref{app:A} for more information.
The algorithm to compute the modified state matrix $\tilde{\mathsf{S}}$ is given below, where $\lfloor y \rfloor$ is the floor function rounding $y$ down to the greatest integer less than or equal to $y$, getBinomialTerm(i,j,p) returns $\binom{k+i}{i} p^i m^k$ and $w$ is the mask weight vector of length $N_V$. 
The implemented C++ code can be found in the supplementary.

\begin{algorithm}
	\SetKwData{State}{State}
	
	\State $\in \mathbb{R}^{N_V \times K}$\;
	\For {$\text{RowsInPascalsTriangle} < \text{maxRow}$}
	{
		\For{$\text{ColumnsInRow} < \text{RowNumber}+1$}
		{
			\For{$\text{VirtualNeuron} < \text{NumberVirtualNeurons}$}
			{
				n = VirtualNeuron\;
				i = RowsInPascalsTriangle\;
				k = ColumnsInRow\;
				\State($\lfloor (n+i+k\nu)/N_V \rfloor$,$n$) += $\text{getBinomialTerm}(i,k,p) \cdot w_{n+i+k\nu \mod N_V}$\;
				\uIf{$\text{getBinomialTerm}(i,k,p) < \sigma$ for all ColumnsInRow}{
					\Return \State\;
				}
			}
		}
	}
	\caption{Calculate modified state matrix}
	\label{alg:s_nl}
\end{algorithm}

\subsection{Direct simulation of the reservoir and memory capacity}

Simulations have been performed in standard C$++$.
For linear algebra calculations, the linear algebra library "Armadillo" \cite{SAN16} was used.
To numerically integrate the delay-differential equations,
a Runge-Kutta fourth-order method was applied, with integration step $\Delta t=0.01$ in dimensionless time units.
First, the system is simulated without reservoir inputs, thus letting transients decay.
After that, a buffer time of 10000 inputs was applied (this is excluded from the training process).
In the training process, $K=75000$ inputs were used to have sufficient statistics. Afterward, the memory capacities $C_{\mathbf{l}}$ of linear recalls were calculated with Eq.~(\ref{eq:mpsi_mem_capacity}), whereby a testing phase is not necessary.
The linear memory capacity $\text{MC}$ was calculated by summing the obtained capacities $C_{\mathbf{l}}$. 
For the piecewise-constant $T$-periodic mask function $g(t)$ independent and identically distributed random numbers between $[0,1]$ were used.

For all simulations, the input strength $\eta$ was fixed to $10^{-3}$. The small input strength was used to guarantee linear answers of the reservoir and, hence, the relevance of the approximation.

A program written in C++ to perform the semi-analytic calculations is given in the supplementary material.

\begin{figure}%
	\centering
	\includegraphics[width=\textwidth]{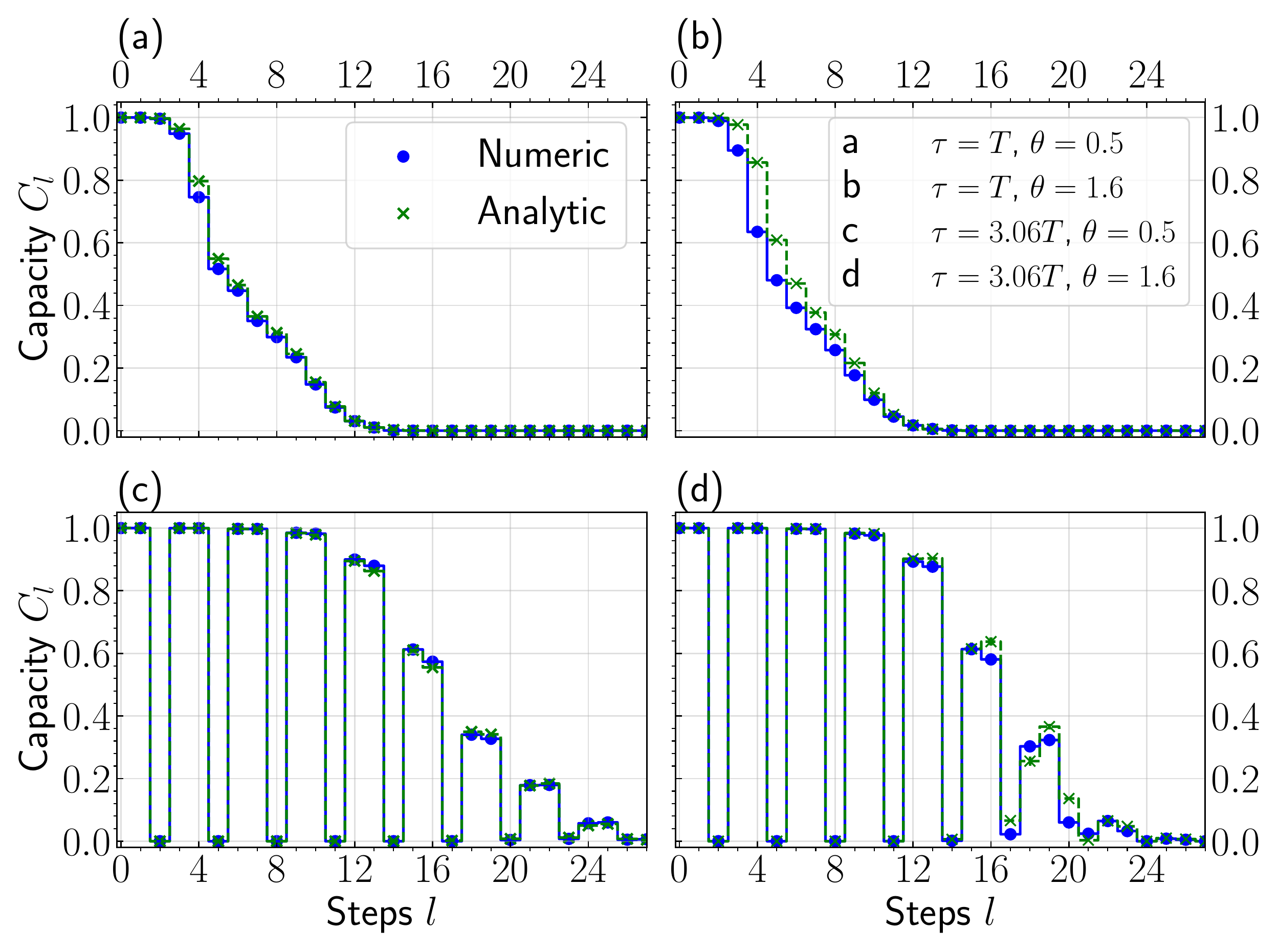}
	\caption{Linear memory capacity $C_l$ computed directly (blue dots) and via MMF by Eq.~\eqref{eq:Cl} (green crosses) for the Stuart-Landau reservoir computer, plotted as a function of the recall steps $l$. (a,c) $\theta=0.5$, (b,d) $\theta=1.6$ and (a,b) $\tau=T$, (c,d) (green) $\tau=3.06T$. The values are averaged over 100 different masks. The parameters for all four setups are $T=80$, $p_{SL}=-0.05$, $\kappa=0.06$, $\eta = 10^{-3}$, and $\gamma = 0.1$.}%
	\label{fig:1d_comparison_1}%
\end{figure}%

\subsection{Comparison of MMF and direct numeric calculations of the memory capacity}

In this section we illustrate the MMF effectiveness. 
First, we show that the MMF provides a very good approximation of MC using the reservoir given by Eq.~(\ref{eq:ST}). The approximation works quite well 
as long as $\theta$ is relatively small. This is fulfilled for typical reservoir computing setups, as one would otherwise lose computation speed.  In the second part, we show how MMF 
provides a universal, system-independent characteristics. For this, we compare MMF with the memory capacities of different reservoirs. Each particular reservoir realization is described by one parameter combination of the MMF. In the last part, we describe how MMF can be computed for reservoirs with unknown dynamical rule. For this, the parameters $a$ and $b$ of the linearization are measured from system's response to a small periodic input. 

Figure \ref{fig:1d_comparison_1} shows the memory recall capacity $C_l$ obtained from direct simulations and compares it with the MMF for 4 different cases of the Stuart-Landau system, given by Eq.~(\ref{eq:ST}).
The exact parameters are given in the caption of Fig.~\ref{fig:1d_comparison_1}. 
The directly simulated results are shown by blue solid lines and blue markers , whereby green dashed lines and green markers show the MMF.
For a small virtual node distance $\theta=0.5$ in Fig. \ref{fig:1d_comparison_1}(a,c), the MMF predicts the linear memory capacity very accurately.
For a higher value of $\theta=1.6$ (Fig. \ref{fig:1d_comparison_1}(b,d)), the accuracy drops, though the results are still accurate for qualitative predictions, and describe the general trend of the system's memory capacity.

The scans in Fig. \ref{fig:1d_comparison_1}(c) and \ref{fig:1d_comparison_1}(d) were done with a higher delay time $\tau=3.06T$, which induces memory gaps \cite{KOE20a}.
Even though the memory capacity has a complex dependency on $l$ at these parameter values, the prediction for the two different virtual node distances $\theta=0.5$ and $\theta=1.6$ is still accurate.

A 2-D parameter plane was simulated in App. \ref{App:2D-Scan} to show that the predictions of the MMF work for arbitrary parameter setups, thus the general predictability of the new scheme is very promising.

Comparing the computation speed of the classical numerical intergration and the new proposed scheme shows an increase of 2 to 3 orders of magnitude, depending on the operation point, the number of training steps $K$ and the value of the clock cycle $T$.
A higher clock cycle $T$ and more training steps $K$ increase the simulation time for the direct numerical integration, whereas the new proposed scheme is independent of that.
If the operation point is close to a bifurcation, the convergence of the new proposed scheme is slower, increasing the computation time needed.
Still, even very close to the bifurcation line, the computation speed is significantly higher (with a factor of about 100) making the MMF a valuable tool.
See App. \ref{app:B}.

\begin{figure}%
	\centering
	\includegraphics[width=\textwidth]{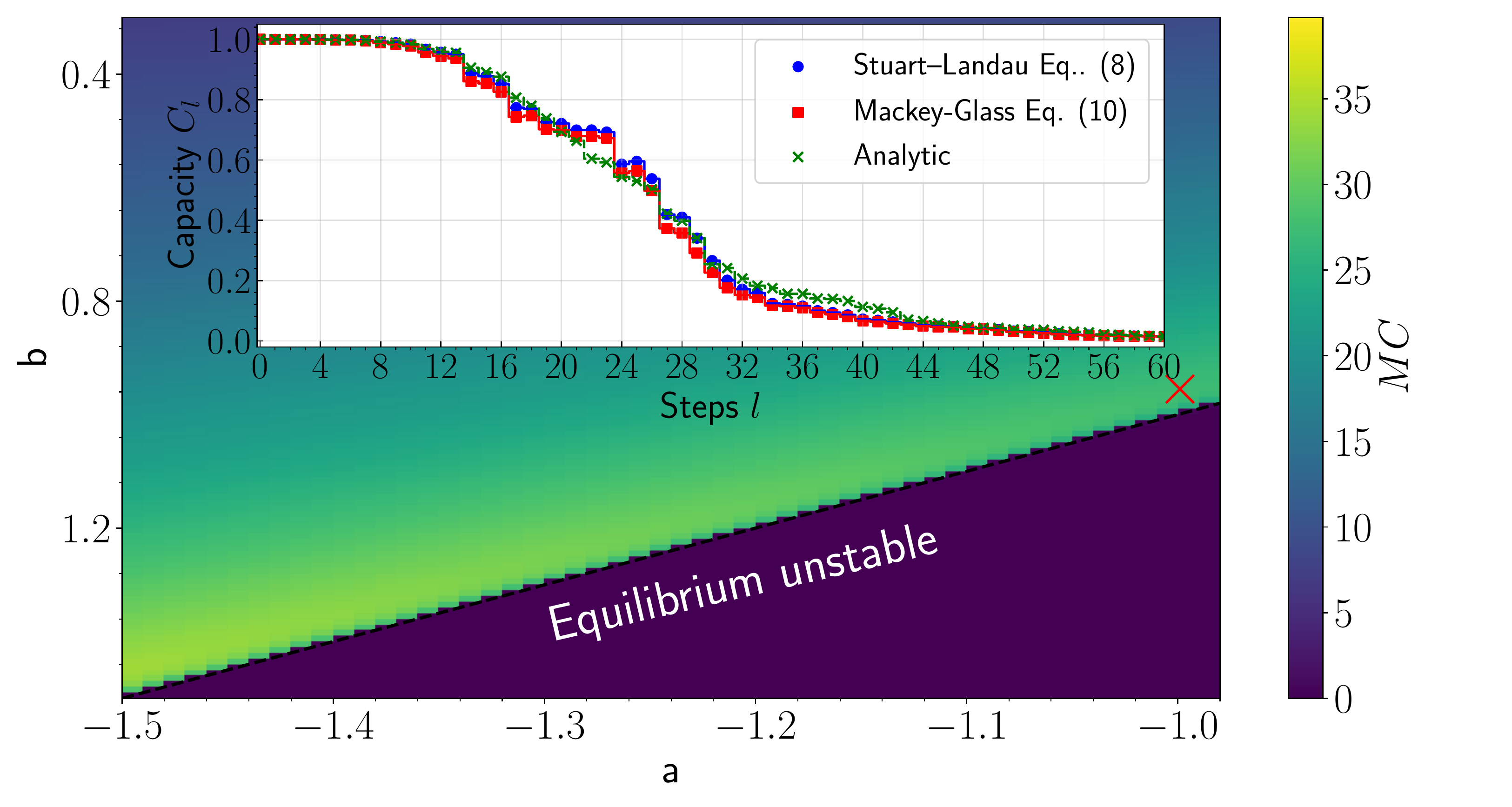}
	\caption{$\text{MC}_{\text{MMF}}$ computed by the MMF in the 2-dimensional parameter plane of the linearization parameters $a$ and $b$, with $N_V=100$, $\tau=72$, $\theta=0.5$. At the edge to instability the performance is highest. The inset shows the MC over the recall steps $l$ of the MMF, the Stuart-Landau and the Mackey-Glass System, at the parameter point indicated at the red cross.
		\label{fig:universality}}
\end{figure}%

\subsection{Universality}

An exciting result that follows from the MMF concept is the possibility to generalize to arbitrary time-delay-based reservoirs.
Every reservoir with a similar linearization should yield similar linear memory capacity.
To illustrate this, we compare the Stuart-Landau reservoir system given by Eq.~(\ref{eq:ST}) and the Mackey-Glass reservoir system given by Eq.~(\ref{eq:MG}).

The inset of Fig.~\ref{fig:universality} illustrates this fact. It shows the capacity to recall the $l$-th step into the past $C_l$ as a function of $l$ for the Stuart-Landau (blue), the Mackey-Glass (red), and the MMF
given by Eq.~\eqref{eq:Cl} (green). Both systems are tuned such that their respective linearization yield the same parameters $a$ and $b$.

From this it follows that it is enough to compute the linearization parameters $a$ and $b$ to predict the MC of any arbitrary delay-based reservoir computer.
The color plot in Fig.~\ref{fig:universality} shows the MMF given by Eq.~\eqref{eq:MC-11} for different parameter values $a$ and $b$. 
A well-performing operation point seems to be the edge to instability, agreeing with the known rule of thumb from the literature.
Any reservoir yielding the same linearization parameters $a$ and $b$ in \eqref{eq:linearization} must possess the corresponding memory capacity as given by Fig.~\ref{fig:universality} for these values of the parameters, as soon as the input is sufficiently small.  

It thus follows that analyzing the Jacobian (linearization given by Eq. (\ref{eq:linearization})) for fixed delay $\tau$, virtual node distance $\theta$, and number of virtual node $N_V$ is sufficient to predict the linear memory capacity of any arbitrary time-delay-based reservoir computer, and this memory capacity is given by MMF via Eqs.~\eqref{eq:MC-11} and \eqref{eq:Cl}.

\subsection{Systems with unknown dynamics; small signal response approach \label{sec:ssr}}

In this chapter, we show an experimentally accessible approach for measuring the parameters $a$ and $b$ for a delay system whose dynamical equations of motion are not known and which can be described by a single variable.
The corresponding linearized dynamical system is  given by
\begin{align}
	\dot{s}(t) = a s(t) + b s(t -\tau) + c I(t).
\end{align}
The goal is to measure $a$ and $b$.
This can be achieved by perturbing the system with a harmonic periodic signal $I(t) = I_0 \sin(\omega t)$.  When this signal is small, we can consider the perturbed linearized system 
\begin{align}
	\label{eq:linear}
	\delta \dot{s}(t) = as(t) + bs(t-\tau) + c I_0 e^{i\omega t},
\end{align}
where the complex form is chosen for simplicity. Due to linearity, the real solution is obtained simply by taking the real part. We consider the case of real $a$ and $b$, which holds always when the reservoir variable is real.

Since the homogeneous solution decays to the stable equilibrium (we assume its exponential stability), the solution of Eq. \eqref{eq:linear} converges to the particular solution, given by
\begin{align}
	x_a(t) = c I_0 H^{-1}(\omega)e^{i\omega t}
\end{align}
with $H^{-1}(\omega) = i\omega - a  - b e^{-i\omega \tau}$.
The ratio of the output to the input amplitude equals to the transfer function
\begin{align}
	\frac{|\text{Output}|}{|\text{Input}|} = \frac{|x_a(t)|}{|c I_0|} = |H^{-1}(\omega)|,
\end{align}
where $|H(\omega)|$ can be measured.

To determine the parameters $a$ and $b$, it is sufficient to measure the transfer function at two frequencies, for example, at  $\omega_R = 2 \pi/\tau$ and $\omega_A = \pi/\tau$. The first frequency is resonant to the delay while the second is in 'anti-phase' to the delay $\tau$. 
It holds 
$$
F(\omega_R) := |H^{-1}(\omega_R)|^2 =  \omega_R^2 + (a+b)^2,
$$
$$
F(\omega_A) := |H^{-1}(\omega_A)|^2 =  \omega_R^2 + (a-b)^2.
$$
From above we can obtain the values for $a$ and $b$
\begin{align}
	a &= -\frac{1}{2} \left( \sqrt{F(\omega_R) - \omega_R^2} + \sqrt{F(\omega_A) - \omega_A^2} \right), \\
	b &= \frac{1}{2} \left( \sqrt{F(\omega_A) - \omega_A^2} -\sqrt{F(\omega_R) - \omega_R^2} \right),
\end{align}
where the values of $F(\omega_A)$ and $F(\omega_R)$ can be obtained experimentally or numerically by perturbing and measuring the response of the reservoir. 

We remark, that the choices of the resonant and anti-phase perturbation frequencies are convenient, but not unique. Clearly, one can perturb at other frequencies to obtain $a$  and $b$. Moreover, the above idea can be generalized to the case of complex-valued parameters $a$ and $b$, whereby more frequencies must be tested. 

The measured values of the parameters $a$ and $b$ for a reservoir with unknown dynamics can be then simply used in MMF by estimating the linear memory capacities. 

\section{Conclusions and discussion}

We have developed a simple and fast method for calculating the linear memory capacity for time-delay-based reservoirs with single-variable dynamics.
The method allows the construction of a modified state matrix whose columns point in the direction of the linear recall steps. 

Our results can be used to predict the reservoir computing setup with high linear memory capacity. 
The nonlinear memory capacity, on the other hand, remains an open question.
In this case, combined setups could be used, where a delay-based reservoir computer includes multiple uncoupled subsystems.
The decoupling ensures that no highly complex dynamical responses destroys the computational performance of the reservoir.
One time-delay-based reservoir computing subsystem can be tuned to low perturbations at the edge of instability to act as a high linear memory kernel.
Increasing the perturbation strength for the other subsystems will ultimately increase the nonlinear responses and thus the nonlinear memory capacity, so that the subsystems with high input strengths take on the role of high nonlinear transformation kernels. 

A teamwork setup is thus recommended, where one or a few subsystems perturbed by small inputs and operated close to instability act as a linear memory kernel. In contrast, other nodes are perturbed more strongly and thus act as highly nonlinear transformation units.
Such a setup should be capable of tackling a wide range of different tasks.
It would be interesting to investigate this in future works.

One of the advantages of the delay-based reservoir, which allows the introduction of the MMF, is that it contains a small number of system parameters while the dynamics remains infinite dimensional. 
In the case of a small input signal and single-variable dynamics, these are only the linearization parameters $a$ and $b$. 
Thus, if the linear memory capacity is computed for all possible values of these two parameters, it covers the case of all possible reservoirs. 
This procedure could be difficult, if not impossible, for network-based reservoirs, where the systems parameters may include, e.g., multiple coupling weights.

\appendix

\section{Derivation of the modified state matrix and reduced formula for memory capacity}
\label{app:A}
Consider a single-variable delay-differential equation, which describes the dynamics of the reservoir
\begin{align}
	\label{eq:DDE}
	\dot s(t) = F(s(t), s(t-\tau), I(t)),
\end{align}
where $I(t)$ stands for an external input. We assume that $s^*$ is an equilibrium of this system, i.e., 
$F(s^*, s^*, 0)=0$, and $I(t)$ is small. 
In the case when the dynamics of \eqref{eq:DDE} takes place near the equilibrium point $s^*$, we can introduce the perturbative ansatz $s(t)=s^*+\delta s(t)$. Then the linearized system for the perturbation $\delta s(t)$  reads
\begin{align}
	\label{eq:ds-linearized}
	\delta \dot{s}(t) = a \delta s(t) +b \delta s(t -\tau) +cI(t),
\end{align}
where $a=\partial_1 F (s^*, s^*, 0)$, $b=\partial_2 F (s^*, s^*, 0)$, and $c=\partial_3 F (s^*, s^*, 0)$. 

Consider $\theta$ to be the temporal node-spacing of the reservoir, which is typically $\theta<\tau$. Then, the equation \eqref{eq:ds-linearized} on any interval $[j\theta,(j+1)\theta]$ can be considered as the simple scalar ODE $\delta \dot{s}(t) = a \delta s(t)$ with the inhomogeneity $b \delta s(t -\tau) + cI(t)$. Moreover, according to the reservoir setup, the input is constant on this interval and equals $I(t)=I_j$. By variation of constants formula, we obtain for the solution of \eqref{eq:ds-linearized} for $t \in [(j-1)\theta,j\theta]$
\begin{align}
	\label{eq:ds111}
	\delta s(t) = -\frac{cI_{j}}{a} (1-e^{a(t-t_{j-1})}) + 
	\delta s(t_{j-1}) e^{a(t-t_{j-1})} \nonumber \\
	+ b \int_{t_{j-1}}^t e^{a(t-\xi)}\delta s(\xi-\tau) d\xi,
\end{align}
where $t_{j-1}=(j-1)\theta$ is the left endpoint of the interval.

\begin{figure}
	\begin{center}
		\begin{tikzpicture}[%
			auto,
			block/.style={
				rectangle,
				draw=blue,
				thick,
				fill=blue!20,
				text width=1.5em,
				align=center,
				rounded corners,
				minimum height=1.2em
			},
			block1/.style={
				rectangle,
				draw=blue,
				thick,
				fill=blue!20,
				text width=2.5em,
				align=center,
				rounded corners,
				minimum height=1.2em
			},
			block2/.style={
				rectangle,
				draw=blue,
				thick,
				fill=blue!20,
				text width=3.5em,
				align=center,
				rounded corners,
				minimum height=1.2em
			},
			block3/.style={
				rectangle,
				draw=red,
				thick,
				fill=blue!20,
				text width=1.5em,
				align=center,
				rounded corners,
				minimum height=1.2em
			},
			block4/.style={
				rectangle,
				draw=red,
				thick,
				fill=blue!20,
				text width=5.0em,
				align=center,
				rounded corners,
				minimum height=1.2em
			},
			block5/.style={
				rectangle,
				draw=red,
				thick,
				fill=blue!20,
				text width=9em,
				align=center,
				rounded corners,
				minimum height=1.2em
			},
			block6/.style={
				rectangle,
				draw=green,
				thick,
				fill=blue!20,
				text width=1em,
				align=center,
				rounded corners,
				minimum height=1em
			},
			block7/.style={
				rectangle,
				draw=green,
				thick,
				fill=blue!20,
				text width=0.7em,
				align=center,
				rounded corners,
				minimum height=1em
			},
			block8/.style={
				rectangle,
				draw=green,
				thick,
				fill=blue!20,
				text width=1em,
				align=center,
				rounded corners,
				minimum height=1em
			},
			line/.style={
				draw,thick,
				-latex',
				shorten >=2pt
			},
			cloud/.style={
				draw=red,
				thick,
				ellipse,
				fill=red!20,
				minimum height=1em
			}
			]
			
			\draw[thick, ->] (-1.4,0) -- (1,0);
			\node[block] at (-2,0) {${s_j}$};
			\node[block6] at (-0.25,0.3) {$\gamma$};
			\node[block3] at (1.5,0) {$ F_{j}$};
			
			\draw[thick, <-] (-2.05,-0.5) -- (-3,-2);
			\draw[thick, <-] (-1.95,-0.5) -- (-1,-2);
			\node[block7] at (-2.8,-1.0) {$p$};
			\node[block8] at (-1.2,-1.0) {$m$};
			
			\node[block1] at (-3,-2.45) {$s_{h-1}$};
			\node[block1] at (-1,-2.45) {$s_{h-\nu}$};
			\draw[thick, ->] (-0.25,-2.45) -- (1.5,-2.45);
			\node[block6] at (0.5,-2.15) {$ \gamma$};
			\node[block4] at (2.6,-2.45) {$ F_{j-1}, F_{j-\nu}$};
			
			\draw[thick, <-] (-3.2,-2.95) -- (-4,-4.45);
			\draw[thick, <-] (-0.75,-2.95) -- (0.0,-4.45);
			\draw[thick, <-] (-3,-2.95) -- (-2.25,-4.45);
			\draw[thick, <-] (-1,-2.95) -- (-1.75,-4.45);
			\node[block7] at (-3.85,-3.5) {$p$};
			\node[block8] at (-2.3,-3.5) {$ m$};
			\node[block8] at (-0.1,-3.5) {$m$};
			\node[block7] at (-1.65,-3.5) {$ p$};

			\node[block1] at (-4,-4.9) {$s_{h-2}$};
			\node[block1] at (0,-4.9) {$s_{h-2\nu}$};
			\node[block2] at (-2,-4.9) {$s_{h-1-\nu}$};
			\draw[thick, ->] (0.75,-4.9) -- (1.5,-4.9);
			\node[block6] at (1.1,-4.6) {$ \gamma$};
			\node[block5] at (3.3,-4.9) {$ F_{j-2},F_{j-1-\nu}, F_{j-2\nu}$};
			
		\end{tikzpicture}
	\end{center}
	\caption[Pscals Trianle]{Pascals Triangle showing the series contributions for $s_j$, given by all the participating equilibria $s^{*}_{j-i-k\nu}$. Blue boxes show participating timesteps, green boxes show multiplications by time propagating factors and red boxes give equilibria contributions for the specific timesteps. Out of convenience, we denote $m= -\frac{b}{a}(1-p)$ and $\gamma= (1 + \frac{b}{a})(1-p)$. $F_j$ denotes the fixpoint factor contributions given by $\gamma \binom{k+i}{i} p^i m^k I_{j-i-k\nu}$ for a specific $i,k$.
	}
	\label{fig:pascals}
\end{figure}
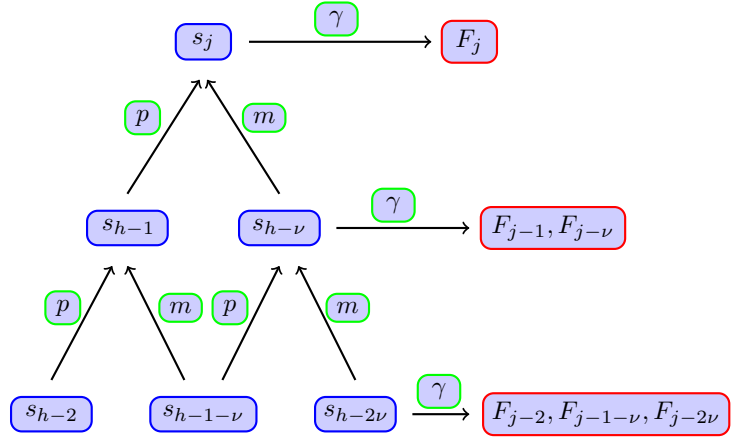

Denoting $\delta s_j(t)=\delta s(t_{j-1}+t)$ to be the function on the interval $[t_{j-1},t_{j-1}+\theta]$, with $t\in(0,\theta)$, we rewrite equation \eqref{eq:ds111} as 
\begin{align}
	\label{eq:ds222}
	\delta s_j(t) = -\frac{cI_{j}}{a} (1-e^{at}) + 
	\delta s_{j-1}(\theta) e^{at} \nonumber \\
	+ b \int_{0}^t e^{a(t-\xi)}\delta s_{j-\nu}(\xi) d\xi,\quad t\in [0,\theta],
\end{align}
where we additionally used the relation $\delta s_{j-1}(\theta)=\delta s_{j}(0)$ and 
$\delta s_{j-\nu}=\delta s_{j}(t-\tau)$, $\nu=\tau/\theta$.
By evaluating Eq.~\eqref{eq:ds222} at $t=\theta$, we obtain
\begin{align}
	\label{eq:ds222-1}
	\delta s_j(\theta) = -\frac{cI_{j}}{a} (1-p) + 
	\delta s_{j-1}(\theta) p \nonumber \\
	+ bp \int_{0}^\theta e^{-a\xi}\delta s_{j-\nu}(\xi) d\xi,\quad p=e^{a\theta}.
\end{align}
Denote $s_j:=s_j(\theta)=s^*+\delta s_j(\theta)$, which is the approximation for state of the reservoir \eqref{eq:DDE} at the virtual nodes $s(j\theta)$. From \eqref{eq:ds222-1}, we obtain
\begin{align}
	\label{eq:sappr}
	s_j = (1 - \frac{b}{a} p) (1-p) s^* 
	- \frac{cI_{j}}{a} (1-p) + 
	p s_{j-1}  \nonumber \\
	+ bp \int_{0}^\theta e^{-a\xi} s_{j-\nu}(\xi)d\xi.
\end{align}
Further, we approximate the integral from Eq.~\eqref{eq:sappr} by assuming $s_{j-\nu}(\xi)\approx s_{j-\nu}(\theta)=s_{j-\nu}$. The approximation holds, in particular, when $\theta$ is small. The obtained expression
\begin{align}
	\label{eq:sappr1}
	s_j = \hat s^* +
	\gamma I_j + 
	p s_{j-1} 
	+ m  s_{j-\nu}
\end{align}
represents a discrete map (coupled map lattice) for approximating the state matrix $\mathsf{S}$. 
Here $\hat s^* = (1 - \frac{b}{a} p) (1-p) s^* $, $m=-(b/a)(1-p)$, and $\gamma = - (c/a) (1-p)$. 
If considering it as a corresponding network with the nodes $s_j$, see e.g. \cite{STE21,Stelzer2020a,HAR17a}, we see that the node $s_j$ is coupled with the two nodes $s_{j-1}$ and $s_{j-\nu}$ in a feed-forward manner with the coupling weights $p$ and $m$, respectively. The schematic representation of such a coupling structure leads to a Pascal's triangle shown in Fig.~\ref{fig:pascals}.
The first row  of  the Pascal's triangle from Fig.~\ref{fig:pascals} shows the dependence on $I_{j}$, which is simply the multiplication by $\gamma$. In the second row, the contributions of $I_{j-1}$ and $I_{j-\nu}$ are shown. To obtain these dependencies explicitly, we insert $s_{j-1}$ and $s_{j-\nu}$ recursively in \eqref{eq:sappr1}:
\begin{align}
	\label{eq:PD1row}
	s_{j}=\left(1+p+m\right)\hat{s}^{*}+\gamma\left(I_{j}+pI_{j-1}+mI_{j-\nu}\right) \nonumber \\
	+p^{2}s_{j-2}+2pms_{j-\nu-1}+m^{2}s_{j-2\nu},
\end{align}
that is, we obtain the terms $\gamma p I_{j-1}$ and $\gamma m I_{j-\nu}$.
To build up further intuition about the dependence of the state matrix on the input, we show here the third level by substituting recursively $s_{j-2}$, $s_{j-\nu-1}$, and $s_{j-2\nu}$ into Eq.~\eqref{eq:PD1row}:
\begin{align}
	\label{eq:PD2row}
	s_{j}= & \left(1+p+p^{2}+2pm+m+m^{2}\right)\hat{s}^{*} \nonumber \\
	& + \gamma\left(I_{j}+pI_{j-1}+mI_{j-\nu}+p^{2}I_{j-2} \right. \nonumber \\
	& \left. +2pmI_{j-\nu-1}+m^{2}I_{j-2\nu}\right) \nonumber \\
	& + p^{3}s_{j-3}+3p^{2}ms_{j-2-\nu}+3pm^{2}s_{j-2\nu-1} + m^{3}s_{j-3\nu}.
\end{align}
To obtain a general recursive formula, we need to split the index in the appearing terms as $j-i-k\nu$ , where $k$ corresponds to the delayed ('right', $m$) and $i$ to the 'left' ($p$) connections in the coupling network in Fig.~\ref{fig:pascals}:
\begin{align}
	\label{eq:PD2general}
	s_{j}= & A_1 s^{*} + 
	\gamma \sum_{i,k=0} \binom{k+i}{i} p^i m^k I_{j-i-k\nu},
\end{align}
where $A_1$ is a constant depending only on $p$ and $m$. For an infinitely long input sequence, the sum in \eqref{eq:PD2general} goes for all $i,k$ from 0 to $\infty$. Practically, the sum is considered for the available data $I_j$.
As a result, the reservoir states $s_j$ are composed of a linear combination of the inputs
with corresponding coefficients given in Eq.~\eqref{eq:PD2general}.

The elements of the state matrix $\mathsf{S}$ used in the reservoir computing setup are 
$$
[\mathsf{S}]_{jn} = \hat s_{j N_V  + n} = s_{j N_V  + n} - 
\left\langle s_{\cdot\, N_V  + n} \right\rangle,
$$
where  $\left\langle s_{\cdot\, N_V  + n} \right\rangle$ is the average over the input intervals
$\left\langle s_{\cdot\, N_V  + n} \right\rangle = 
(1/K) \sum_{j'=1}^{K}  s_{j' N_V  + n}$. Here and later, the dot denotes the index, over which the averaging is performed.
Taking into account Eq.~\eqref{eq:PD2general},  we obtain
\begin{align}
	\label{eq:PD2general-1}
	\hat s_{j N_V  + n} = &\gamma \sum_{i,k=0} \binom{k+i}{i} p^i m^k \nonumber \\
	&\times \left(
	I_{j N_V  + n -i-k\nu} - 
	\left\langle I_{\cdot \, N_V  + n -i-k\nu} \right\rangle
	\right)
	.  
\end{align}
The input $I_k$ of the reservoir computer is given by the discrete input sequence $\mathbf{u}$ multiplied by the input weights:
$$
I_{k}=u_{\left\lfloor k/N_{V}\right\rfloor }
w_{k\, \mathrm{ mod } \, N_{V}}.
$$

Therefore, we obtain 
\begin{align}
&\left\langle I_{\cdot\,N_{V}+n-i-k\nu}\right\rangle = \nonumber \\
&w_{n-i-k\nu\,\mathrm{mod}\,N_{V}}\left\langle u_{\cdot\, + \left\lfloor \left(n-i-k\nu\right)/N_{V}\right\rfloor }\right\rangle \approx 0,
\end{align}
since $\mathbf{u}$ has zero mean.  
Hence, we have for the elements of the state matrix 
\begin{equation}
	\label{eq:PD2general-2}
	\hat s_{j N_V  + n} = \gamma \sum_{i,k=0} \binom{k+i}{i} p^i m^k 
	I_{j N_V  + n -i-k\nu}
	.  
\end{equation}
Correspondingly, the elements of the covariance matrix $\mathsf{S}^T\mathsf{S}$ from
\eqref{eq:mpsi_mem_capacity} are 
\begin{align}
	\label{eq:apt1}
	\left[\mathsf{S}^{T}\mathsf{S}\right]_{nn'}
	=
	\sum_{l}[\mathsf{S}]_{ln}[\mathsf{S}]_{ln'}
	=
	\sum_{l}\hat s_{lN_{V}+n} \hat s_{lN_{V}+n'},
\end{align} 
and they describe the covariance of the virtual node $n$ with virtual node $n'$ over all clock cycles $l$.

By substituting \eqref{eq:PD2general-2} into \eqref{eq:apt1}, we obtain
\begin{align}
	\left[\mathsf{S}^{T}\mathsf{S}\right]_{nn'} =
	\gamma^{2}\sum_{l}\left(\sum_{k,i}\binom{k+i}{i}p^{i}m^{k}I_{lN_{V}+n-i-k\nu}\right) \nonumber \\
	\left(\sum_{k',i'}\binom{k'+i'}{i'}p^{i'}m^{k'}I_{lN_{V}+n'-i'-k'\nu}\right).
\end{align}

One can show that the elements from $\left[\mathsf{S}^{T}\mathsf{S}\right]_{nn'}$ containing mixed terms of the form $u_{i}u_j$, $i\ne j$ can be approximated by zero since the random variable $u_j$ is independently distributed with zero mean. The only nonzero second moment, which matters in $\left[\mathsf{S}^{T}\mathsf{S}\right]_{nn'}$, is the mean square 
\begin{align}
\sum_{k=0}^\infty u_k^2 = \frac{1}{3}.
\end{align} 
Hence, for further simplification of $\left[\mathsf{S}^{T}\mathsf{S}\right]_{nn'}$, we keep only terms of the form $u_k^2$. The following calculations formalize these arguments. 
\begin{strip}
\begin{align}
&\left[\mathsf{S}^{T}\mathsf{S}\right]_{nn'}= \gamma^{2}\sum_{l} \left(\sum_{k,i}\binom{k+i}{i}p^{i}m^{k}u_{l+\left\lfloor \left(n-i-k\nu\right)/N_{V}\right\rfloor }w_{\left(n-i-k\nu\right)\,\mathrm{mod}\, N_{V}}\right) \nonumber \\
&\times\left(\sum_{k',i'}\binom{k'+i'}{i'}p^{i'}m^{k'}u_{l+\left\lfloor \left(n'-i'-k'\nu\right)/N_{V}\right\rfloor }w_{\left(n'-i'-k'\nu\right)\,\mathrm{mod}\, N_{V}}\right) \nonumber \\
&\approx\gamma^{2}\sum_{j=1}\left(\sum_*\binom{k+i}{i}p^{i}m^{k}\binom{k'+i'}{i'}p^{i'}m^{k'}w_{\left(n-i-k\nu\right)\,\text{mod}\,N_{V}}w_{\left(n'-i'-k'\nu\right)\,\text{mod}\,N_{V}}\sum_{l}u_{l+j-1}^{2}\right)  \nonumber \\
	\label{eq:appr}
	&\approx
	\gamma^{2}\frac{1}{3}
	\sum_{j=1}
	\left(\sum_*\binom{k+i}{i}p^{i}m^{k}\binom{k'+i'}{i'}p^{i'}m^{k'}w_{\left(n-i-k\nu\right)\,\text{mod}\,N_{V}}w_{\left(n'-i'-k'\nu\right)\,\text{mod}\,N_{V}}\right),
\end{align}
\end{strip}
where the second summation range (*) is taken over all values of $k,i,k',i'$ such that 
$n+N_{V} j \le i+k\nu < n+N_{V} (j+1)$ and 
$n'+ N_{V} j \le i'+k'\nu < n'+ N_{V} (j+1)$.

The obtained expression \eqref{eq:appr} does not depend on the sequence $u_j$ and hence, provides a significant simplification for calculating the covariance matrix. We may further notice that the same covariance \eqref{eq:appr} can be obtained by defining the modified state matrix 
$\tilde{\mathsf{S}} = \tilde{s}_{jn}$, where $j$ is the the $j-$th interval of the shifted input (the $j$-th recall) and $n$ the $n$-th virtual node.
$\tilde{s}_{jn}$ is given by the sum over all combinations $i,k$, that fall into the same shifted input interval $j$, i.e.
\begin{equation}
	\tilde{s}_{jn} = 
	\frac{\gamma}{\sqrt{3}}
	\sum_{n + N_{V} j \le i+k\nu}^{i+k\nu < n + N_{V} (j+1)}
	\binom{k+i}{i}p^{i}m^{k}
	w_{\left(n-i-k\nu\right)\,\text{mod}\,N_{V}}.
	\label{eq:stislde}
\end{equation}
This is our main result, because Eq.~\eqref{eq:stislde} defines the modified state matrix $\tilde{\mathsf{S}}$ from which all capacities $C_{l}$ are derivable. More specifically, we have shown
\begin{align}
	\mathsf{S}^T\mathsf{S} \approx
	\tilde{\mathsf{S}}^T \tilde{\mathsf{S}}.
\end{align}

Further, for the $l$-th recall, where the target  is the shifted input sequence
$
\mathbf{\hat y}=\{u_{j-l}\}_{j=1}^\infty$, we have 
\begin{strip}
	\begin{align}
&\left[\mathsf{S}^{T}\hat{\boldsymbol{y}}_{l}\right]_{n}
=
\sum_{j}[\mathsf{S}]_{jn}\left[\hat{\boldsymbol{y}}_{l}\right]_{j}
=
\sum_{j} \hat s_{jN_{V}+n}u_{j-l}
\nonumber \\
&=\sum_{j}
\left(
\gamma\sum_{k,i}\binom{k+i}{i}p^{i}m^{k}u_{j+\left\lfloor (n-i-k\nu)/N_{V}\right\rfloor }w_{n-i-k\nu
	\,\mathrm{mod}\, N_{V}}
\right)
u_{j-l}
 \nonumber \\
&=\gamma\sum_{k,i}\binom{k+i}{i}p^{i}m^{k}w_{n-i-k\nu\,\mathrm{mod}\, N_{V}}\sum_{j}u_{j+\left\lfloor (n-i-k\nu)/N_{V}\right\rfloor }u_{j-l}
\nonumber \\
&\approx\frac{\gamma}{3}\sum_{n+(l-1)N_{V}<i+k\nu}^{i+k\nu\le n+lN_{V}}\binom{k+i}{i}p^{i}m^{k}w_{n-i-k\nu
	\,\mathrm{mod}\, N_{V}}=\frac{1}{\sqrt{3}}\tilde{s}_{ln},
\end{align}
	\end{strip}

therefore, it holds
$$
\mathsf{S}^{T}\hat{\boldsymbol{y}}_{l} \approx
\frac{1}{\sqrt{3}} \mathsf{\tilde S}_l, 
$$
where $\mathsf{\tilde S}_l$ is the $l$-th row of $\mathsf{\tilde S}$.
Further, we notice that 
$$
\hat{\boldsymbol{y}}_{l}^{T} \mathsf{S} =
\left( \mathsf{S}^{T}\hat{\boldsymbol{y}}_{l} \right)^T
\approx
\frac{1}{\sqrt{3}} \mathsf{\tilde S}^T_l, 
$$
and $\| \hat{\boldsymbol{y}}_{l} \|^2 \approx 1/3$. As a result, taking into account the definition of the memory capacity \eqref{eq:mpsi_mem_capacity}, we obtain the approximation for the 
capacity of the $l$-th recall $C_l$ by 
\begin{align}
	C_l \approx \tilde{\mathsf{S}}^T_l \left(\tilde{\mathsf{S}}^T \tilde{\mathsf{S}} + \lambda \mathrm{I}\right)^{-1} \tilde{\mathsf{S}}_l.
\end{align}
The results can be understood in such a way, that we constructed the modified state matrix $\tilde{\mathsf{S}}$, such that every column has entries in the statistical direction of the $l$-th shifted input recall.

The full linear memory capacity is then given by the trace
\begin{align}
	\text{MC}_{\text{MMF}} = \text{tr}\left(\tilde{\mathsf{S}}^T \left(\tilde{\mathsf{S}}^T \tilde{\mathsf{S}} + \lambda \mathrm{I} \right)^{-1} \tilde{\mathsf{S}}\right).
\end{align}

\section{Computation Time}
\label{app:B}

\begin{figure}%
	\centering
	\includegraphics[width=0.9\textwidth]{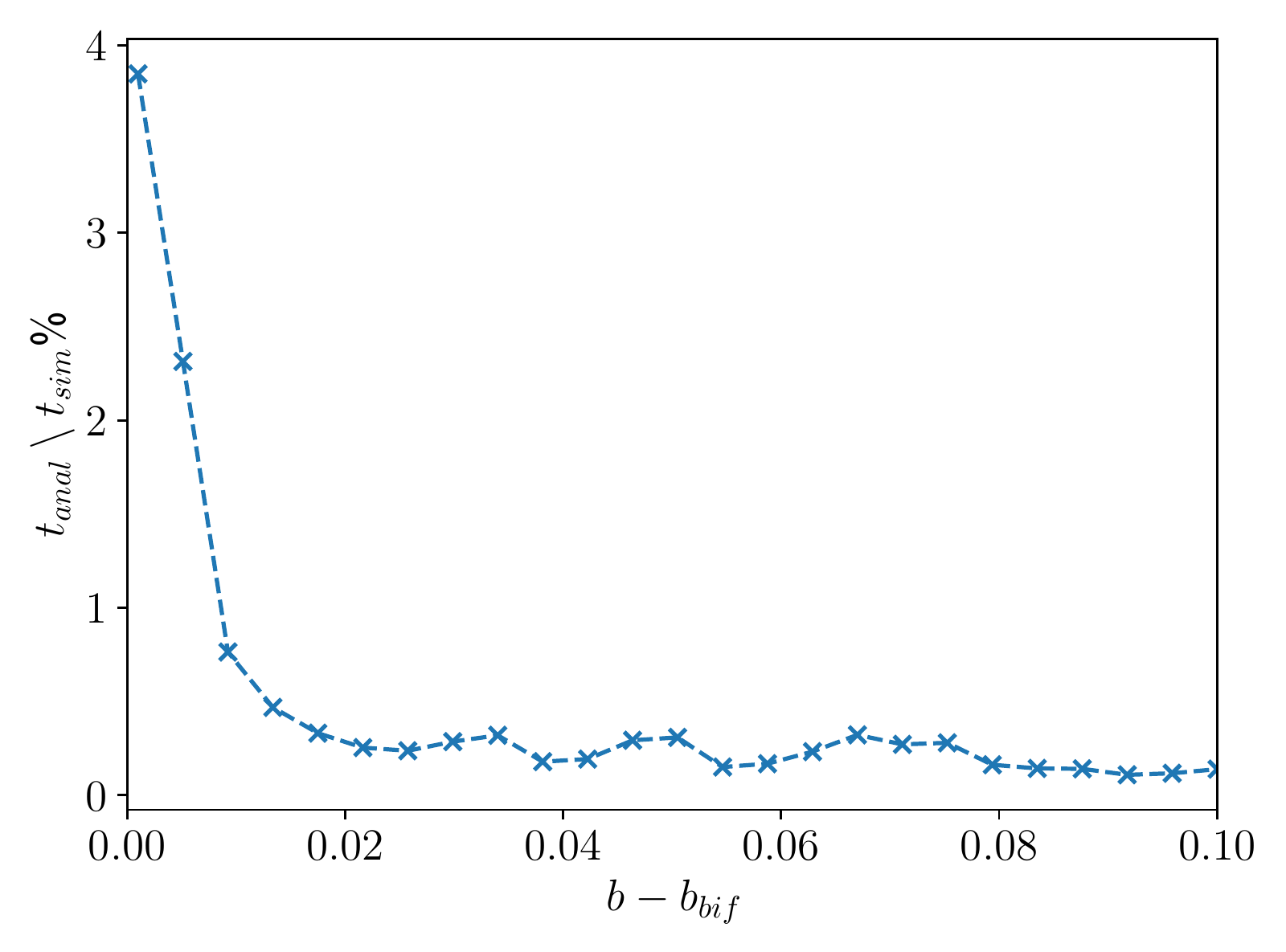}
	\caption{Computation speed of the full simulated system and the analytic approach for a runge-kutta-4 method of timestep $dt=0.01$, a buffer sequence of $10000$ clock cyclces and training samples of $50000$. The analytic approach computed all values in pascals triangle up to $10^{-6} \cdot \max_{s} (\mathsf{S})$. Parameters are, $\tau=141$, $T=100$, $N_V=100$ (i.e. $\theta=1$, $a=-0.503$.} 
	\label{fig:app_comparison_time}%
\end{figure}

To compare the computation speed of the full numerically simulated differential equation and our new analytic approach, we simulated both systems. 
The full system with a time-step of $dt=0.01$, buffer samples of $10000$, i.e. that 10000 clock cycles were simulated and discarded, and $50000$ training samples to get high accuracy on the memory capacity.
The analytic program was calculated until all values in a row in pascals triangle were below $10^{-6} \cdot \max_{s} (\mathsf{S})$.
We compared the simulation speeds of both approaches on a parameter linescan of the linearization parameter $b$, scanning from values close to the bifurcation value $b_{bif}$ in which the linearized system destabilizes up to values of about $0.1$ greater than the bifurcation value $b_{bif}$.
We show the percentage of the simulation time for the analytic approach $t_{anal}$ in comparrison to the simulation time $t_{sim}$, i.e. $t_{anal}\setminus t_{sim} \%$ in Fig.\,\ref{fig:app_comparison_time}.
We see that close to the bifurcation the analytic approach increases in computation time.
This comes from the fact, that the convergence of pascals triangle close to the bifurcation is slower.
Still, the simulation time is at maximum $4\%$ of the fully simulated system, showing at least a $25$-fold increase in computation speed.

\section{Range of Approximation}
\label{app:C}

\begin{figure}%
	\centering
	\includegraphics[width=0.9\textwidth]{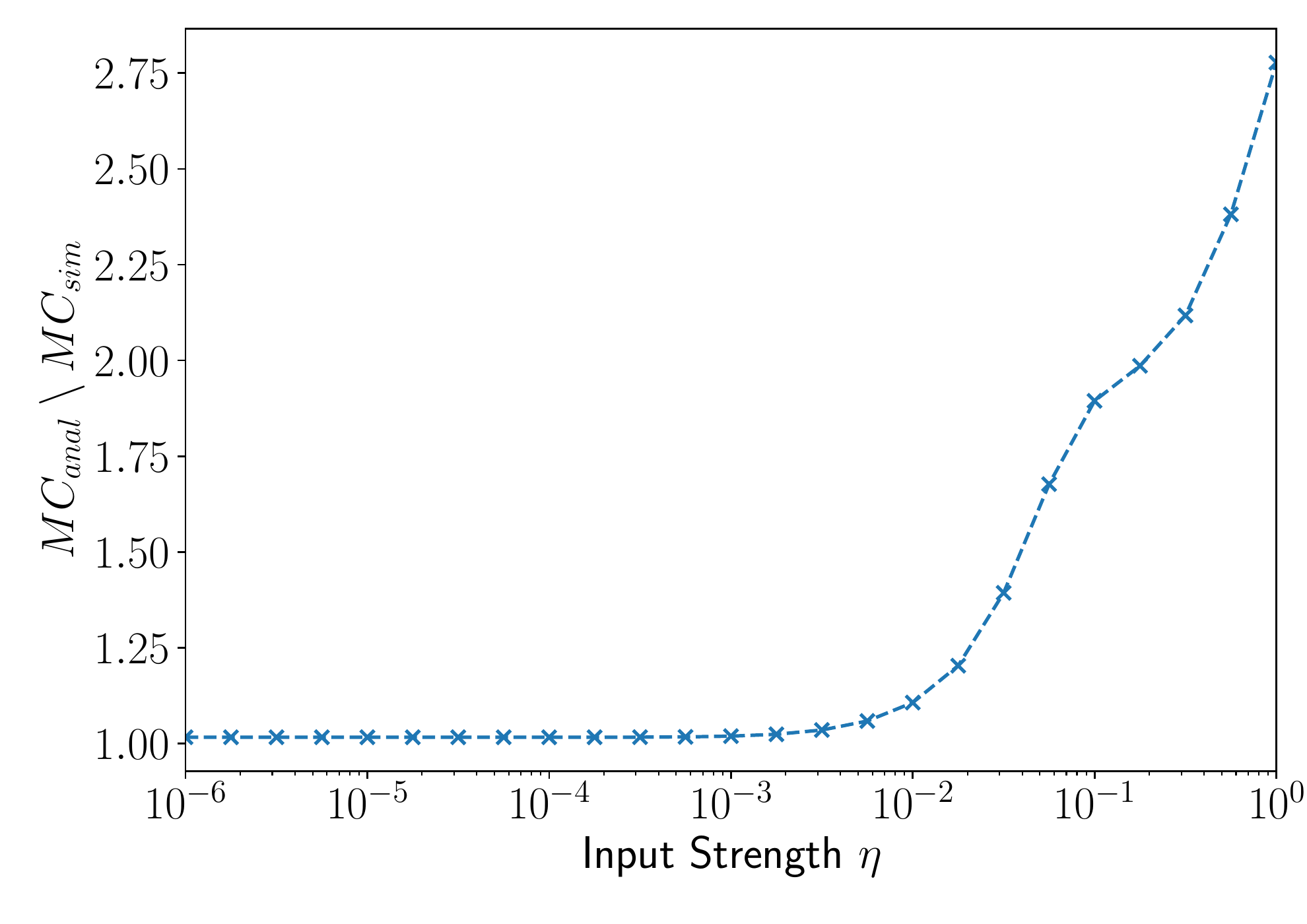}
	\caption{Comparison of memory capacity $MC$ of the fully simulated system $MC_{sim}$ and the analytic approach $MC_{anal}$ over the input strength $eta$. The system was simulated with runge-kutta-4 method of timestep $dt=0.01$, a buffer sequence of $10000$ clock cyclces and training samples of $50000$. The analytic approach computed all values in pascals triangle up to $10^{-6} \cdot \max_{s} (\mathsf{S})$. Parameters are, $\tau=141$, $T=100$, $N_V=100$ (i.e. $\theta=1$, $a=-0.503$, $b=0.201$.} 
	\label{fig:app_comparison_eta}%
\end{figure}

We would like to show the range of approximation for the new analytic approach by computing the memory capacity $MC$ of the fully simulated system $MC_{sim}$ and the analytic approach $MC_{anal}$
by showing the relative memory capacity of the analytic approach to the full system, i.e. $MC_{anal}\setminus MC_{sim}$.
The results are shown in Fig.\,\ref{fig:app_comparison_eta} plotted over the input strength $\eta$ for six magnitudes of order.
A result close to $1$ indicates a good agreement of the simulation and the analytic approach.
For high input strengths, starting at around $\eta=10^{-2}$, the analytic approach overestimates the real memory capacity, because high values of $\eta$ induce nonlinear answers in the system and thus increase the nonlinear transformations of the reservoir in exchange for linear memory. See \cite{DAM12,KOE20a,KOE21} for more information on that effect.

\section{$\chi^2_k$ Estimation}
We give a short insight into the $\chi^2_k$ estimation introduced in \cite{DAM12}.
When calculating capacities $C_{\mathbf{l}}$, all below a fixed value $r^{*}$ were excluded because of finite statistics, where $r^{*}$ is given by the following relation. $\text{CFD}(\chi^2(N_V,r^{*}))$ is the cummulative distribution function of the $\chi^2$ function and $r^{*}$ is chosen such that $1 - \text{CFD}(\chi^2(N_V,r^{*}))$ yields a probability $p_{\chi^{2}}=10^{-6}$, i.e. the probability of a capacity having a value greater than $r^{*}$ even though with infinite statistics ($K \xrightarrow[]{} \infty$), it would have a value less than $r^{*}$.
$\chi^2$ is the probability density function of the sum of squared independent, standard normal random variables
\begin{align}
	\chi^2_k = \sum_{i=1}^{k} Z_i^2
\end{align}
See \cite{DAM12} for more information.

\section{Broader Parameter Range Check}
\label{App:2D-Scan}

\begin{figure}%
	\centering
	\includegraphics[width=\textwidth]{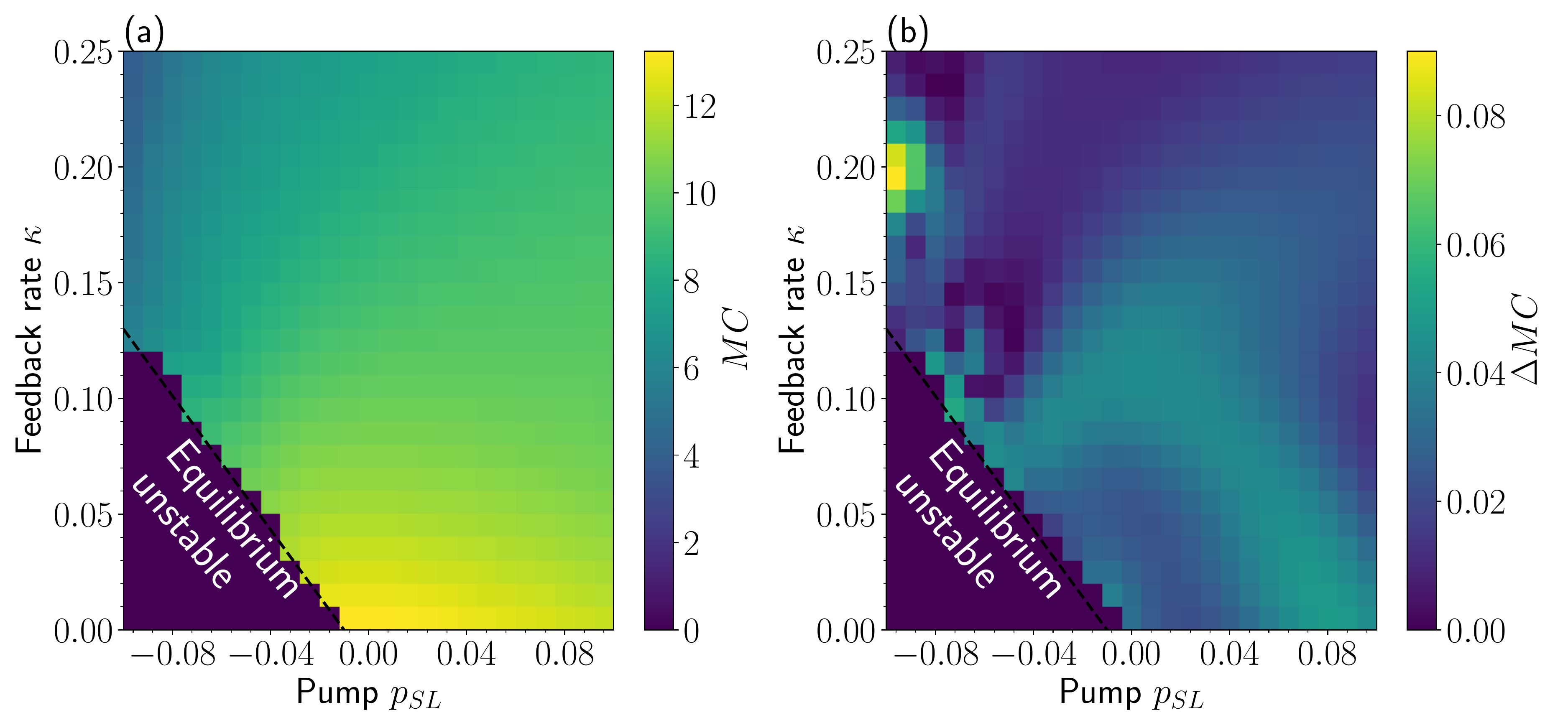}
	\caption{
		Two-parameter characterisation of the memory capacity of system (\ref{eq:ST}) with respect the pump $p_{SL}$ and feedback rate $\kappa$. (a) Total linear memory capacity of the directly simulated  $\text{MC}_\text{direct}$. (b) Relative difference $\Delta \text{MC}$ of the MMF $\text{MC}_\text{MMF}$ and the directly simulated $\text{MC}_\text{direct}$ value. The black dashed line shows the threshold of stabilization of the non-trivial equilibrium. The RV coefficient is $\text{RV}(\text{MC}_\text{direct},\text{MC}_\text{MMF})= 0.99925$. The parameters are $N_V=100$, $T=100$ (corresponding to $\theta=1$), $\tau = 1.41T$, $\eta = 10^{-3}$, and $\gamma = 0.1$.}%
	\label{fig:2d_comparison_1}%
\end{figure}

A 2-parameter characterisation of the memory capacity of the Stuart-Landau system (\ref{eq:ST}) is shown in Fig.~\ref{fig:2d_comparison_1}. 
The parameter space is spanned by the pump $p_{SL}$ and the feedback rate $\kappa$.
Figure \ref{fig:2d_comparison_1}(a) shows the linear memory capacity while Fig.~\ref{fig:2d_comparison_1}(b) shows the relative difference $\Delta \text{MC}$ of the MMF and the direct numerics:
\begin{align}
	\Delta \text{MC} = \frac{|\text{MC}_{\text{MMF}} - \text{MC}_\text{direct}|}{\text{MC}_\text{direct}}.
\end{align}
Small relative differences of up to $0.08$ are seen for the simulations presented here for $\theta=1$.
One has to remember that reservoir computing is usually done with very small $\theta$.
The work \cite{JAE01} introduced a rough estimate of the optimal value for $\theta$ as $\theta \approx 0.2 \lambda_\text{ans}$, where $\lambda_\text{ans}$ is the linear answer timescale of the system. 
In the case of the Stuart-Landau system, this is given by $\lambda_\text{ans} = -2 p_{SL} - 3 \kappa$.
For the parameter space in Fig \ref{fig:2d_comparison_1}, the highest value of the linear answer timescale $\lambda_\text{ans}=-0.95$, thus $\theta$ is by a factor of $5$ bigger than the proposed value given in \cite{JAE01} for optimal virtual node distance.
In our approximation, we assume a constant state value on one $\theta$-interval, thus $\theta=1$ is a very high value, which is one of the reasons for the deviations.

To underline that the MMF (\ref{eq:MC-11}) gives a reasonable estimation of the memory capacity, we also calculated the 2-dimensional correlation coefficient $\text{RV}(X,Y)$ between the directly simulated total linear memory capacity $\text{MC}_\text{direct}$ and the linear memory capacity $\text{MC}_\text{MMF}$ given by MMF in the 2-dimensional plane of the pump $p_{SL}$ and the feedback rate $\kappa$ parameters.
$\text{RV}(X,Y)$ is the generalization of the squared Pearson coefficient for two dimensions and is calculated via:
\begin{align}
	\text{RV}(X,Y) &= \frac{\text{COVV}(X,Y)}{\sqrt{\text{VAV}(X) \text{VAV}(Y)}}
\end{align}
with
\begin{align}
	\Sigma_{XY} &= E(X^TY) \\
	\text{COVV}(X,Y) &= \text{tr}(\Sigma_{XY} \Sigma_{YX}) \\
	\text{VAV}(X) &= \text{tr}(\Sigma^{2}_{XX}),
\end{align}
Here, $E()$ is the expectation value, $\Sigma_{XY}$ denotes the centered covariance matrix of the matrices $X$ and $Y$, $\text{COVV}(X,Y)$ denotes the trace of the matrix multiplication of $\Sigma_{XY} \Sigma_{YX}$ and $\text{VAV}(X)$ the trace of the matrix multiplication $\Sigma^2_{XX}$.
Calculating RV over the parameter range shown in Fig.~\ref{fig:2d_comparison_1} yields a value of $RV(\text{MC}_\text{direct},\text{MC}_\text{MMF}) \approx 0.99925$.
The correlation is close to the maximum of $1$, allowing us to make accurate predictions of high-performing reservoirs with the MMF.

\section*{Acknowledgment}
The authors thank David Hering, Lina Jaurigue, and Joscha Matysiak for fruitful discussions.
This study was funded by the "Deutsche Forschungsgemeinschaft" (DFG) in the framework of SFB910 and project 411803875.

\bibliographystyle{ieeetr}
\bibliography{ref.bib}

\end{document}